\documentclass[amsmath,amssymb,showpacs,showkeywords,twocolumn]{revtex4}
\usepackage[dvips]{graphicx,color}
\usepackage{times}
\usepackage{mathrsfs}
\usepackage{amsmath}
\usepackage{graphicx}
\usepackage{epsfig}
\usepackage{dcolumn}
\usepackage{bm}
\usepackage[utf8x]{inputenc}
\DeclareUnicodeCharacter{2212}{-}
\begin{document}

\title{Nonlinearity in spin dynamics of frustrated Kagom\'e lattice system under harmonic perturbation}
\author{Saumen Acharjee\footnote{saumenacharjee@dibru.ac.in},
Arindam Boruah\footnote{arindamboruah@dibru.ac.in}, Reeta Devi\footnote{reetadevi@dibru.ac.in} and  Nimisha Dutta\footnote{nimishadutta@dibru.ac.in}}
\affiliation{Department of Physics, Dibrugarh University, Dibrugarh 786 004, 
Assam, India}

\begin{abstract}
In this study, we investigate the spin dynamics of a frustrated Kagom\'e lattice system, focusing on the nonlinearity of spin oscillations induced by a harmonic magnetic field under varying strengths of Dzyaloshinskii-Moriya interaction (DMI), exchange field, and anisotropy energy. We have utilized Poincar\'e Surface Sections (PSS) and Power Spectra (PS) for different DMI and anisotropy energy to study the spin dynamics. Our findings reveal that when the DMI strength, external field, anisotropy, and applied magnetic field are weak, the oscillations are quasi-periodic, mostly dominated by the exchange field. With the increase in the DMI strength, the oscillation of the system becomes highly aperiodic. Strong anisotropy tends to induce periodic oscillations, but increasing DMI eventually leads to chaotic behaviour. Additionally, the external magnetic field destabilizes the periodicity of oscillations in systems with weak easy-axis anisotropy, but the systems with strong anisotropy, the oscillations remain unaffected by the external field's strength. Our analysis of magnon dispersion and magnetic resonance (MR) spectra reveals multiple resonance peaks at higher DMI strengths, indicating a complex interplay between spin wave excitation and system parameters. These results underscore the importance of understanding the inherent DMI and anisotropy in the Kagom\'e lattice during fabrication for various applications. Moreover, our comprehensive analysis of spin dynamics in a Kagom\'e lattice system demonstrates a clear transition from quasi-periodic to chaotic oscillations with the increase in DMI strength.

\end{abstract}

\maketitle

\section{Introduction}
 Frustrated magnetic systems have garnered significant attention in the scientific community due to their intriguing and often unexpected properties that challenge conventional paradigms of magnetic ordering. They exhibit unconventional ground states and display exotic magnetic excitation and quasi-particles \cite{ghimire, mielke}. Thus, it may behave unconventionally and contribute to unique magnetic and thermal properties. In general, frustration in a magnetic system arises when the interactions within a given system cannot be simultaneously minimized, leading to a considerable multitude of low-energy configurations characterized by an approximate degeneracy that scales exponentially with the size of the system \cite{khatami, saha, farhan, hog}. Moreover, these states can also include spin liquids \cite{han, yan, nayak}, where magnetic moments remain disordered even at very low temperatures. Among the various frustrated geometries, the Kagom\'e lattice holds particular interest because of its significantly smaller energy gaps than the other geometries, with an exponentially large number of singlet states below the lowest triplet state \cite{han, mendels, li, zorko, du, yin}. Conversely, experiments conducted on materials featuring nearly decoupled Kagom\'e layers, exemplified by Herbertsmithite, reveal the absence of magnetic order even at extremely low temperatures and a lack of discernible evidence for a spin gap. Nevertheless, when the study of these materials is delving into these materials, the magnetic anisotropy and the Dzyaloshinskii - Moria interaction (DMI) \cite{dmi1, dmi2} plays a significant role in dictating the nature of the ground state. 
 
 The DMI is an antisymmetric exchange interaction emerging in systems having no inversion symmetry and promotes chiral magnetic phases such as skyrmions \cite{hog, liu1, chen, petrova, lee, drozdov} in two dimensional (2D) lattices. These unique magnetic textures have attracted significant attention due to their potential applications in spintronics and information storage \cite{wiesendanger, zhang, marrows, dupe, zhou, fert, ishida, zhang2, tokura}. DMI can facilitate the controlled movement and manipulation of domain walls \cite{je, ajejas, brandao, wang, dejong}. Moreover, the nontrivial spin textures induced by DMI leading to a transverse deflection of electrons as they move through the material. Thus, it hosts the topological Hall effect and also leads to a modification in the spin wave spectrum that can be probed experimentally \cite{ren, gao, huang, wang1}. So, the role of DMI in spin dynamics and excitation spectrum of 2D lattice system is too inherent.
 The Kagom\'e lattice, with its inherent frustration from competing interactions, tends to resist conventional magnetic ordering. The DMI introduces non-collinear spin arrangements, favouring the formation of chiral magnetic order. Thus understanding the spin dynamics in the Kagom\'e lattice structure and its interplay with DMI offers a rich playground for exploring novel magnetic phenomena and opens avenues for potential applications in spintronics and magnetic technology.

 The study of spin dynamics is of paramount importance, as it forms the cornerstone for understanding the behaviour of magnetic materials, providing essential insights into their fundamental properties and enabling the design of advanced technologies \cite{acharjee1, acharjee2, acharjee3, acharjee4, acharjee5,liu}.  Over the years, the time evolution of magnetic moments unveiled the intricate interplay of magnetic interactions by exploring diverse magnetic phases in different magnetic systems. This understanding is crucial for developing magnetic materials with tailored functionalities, including applications in information storage, spintronics, and magnetic sensors \cite{wiesendanger, zhang, marrows, dupe, zhou, fert, ishida, zhang2, tokura, gottscholl, khan, yan1, wong}. The deformed Kagom\'e lattice, characterized by its geometric frustration arising from competing interactions within an array of corner-sharing triangles, serves as a fascinating platform for investigating unconventional magnetic behaviour \cite{mendels, li, zorko, du, yin}. It is due to its ability to foster exotic magnetic ground states such as spin liquids, thereby introducing a complex tapestry of interactions among neighbouring spins. Moreover, the Kagom\'e system with anisotropy \cite{du, yin}, becomes even more nuanced, giving rise to intriguing phenomena that challenge the conventional understanding of magnetic systems. This paper delves into the intriguing dynamics of the frustrated spins within an anisotropic Kagom\'e lattice system in the presence of DMI, induced by harmonic perturbation. The basic objective of this work is to understand the role of magnetic anisotropy, DMI in spin dynamics of Kagom\'e system under harmonic perturbation. 

The paper is organized as follows. Section II outlines the minimal theoretical model of the proposed frustrated Kagom\'e system and derives the explicit form of the Hamiltonian. In Section III, we delve into the study of the Poincar\'e Surface Section (PSS), power spectra, and magnetic resonance considering the effect of anisotropy and DMI under harmonic perturbation. This analysis aims to comprehend the nature of the oscillation and stability of spins within the proposed system. Additionally, we explore the finite time Largest Lyapunov exponent (LLE) to investigate the chaos in this system.  Section V concludes our work with a concise summary.
 
\section{Theoretical framework}
\subsection{Spin Dynamics}
\begin{figure}[hbt]
\centerline
\centerline{ 
\includegraphics[scale=0.58]{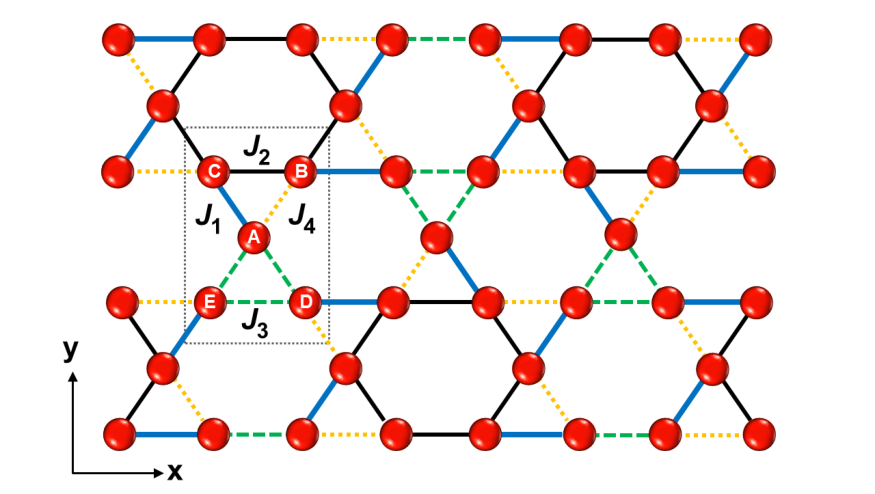}
\vspace{0.1cm}
}
\caption{Schematic representation of two-dimensional anisotropic Kagom\'e lattice with the exchange interactions $J_1$, $J_2$, $J_3$ and $J_4$ between the spins at the sub-lattice  ABCDE. The proposed unit cell with lattice sites ABCDE is denoted by the dotted rectangle. The spin moments at the lattice site A, B and C are denoted by $\mathbf{S}_\text{A}$, $\mathbf{S}_\text{B}$ and $\mathbf{S}_\text{C}$ respectively while the moments at the D and E are chosen to be (0,0,1) for simplicity.}
\label{fig1}
\end{figure}

The schematic representation of two dimensional anisotropic
Kagom\'e lattice with exchange interactions, $J_1$, $J_2$, $J_3$ and $J_4$ is shown in Fig. \ref{fig1}. We consider the lattice sites A, B and C for our reference with the onsite magnetic moment $\mathbf{S}_\text{A}$, $\mathbf{S}_\text{B}$ and $\mathbf{S}_\text{C}$ respectively. The Hamiltonian of the deformed Kagom\'e lattice under harmonic perturbation can be expressed as
\begin{multline}
\label{eq1}
\mathcal{H} = J_\alpha \sum_{\left<i,j\right>} \mathbf{S}_i.\mathbf{S}_j + K \sum_i  (\mathbf{S}_i.\hat{e}_z)^2 + \sum_{ij} \mathbf{D}_{ij}^\text{eff}.\lbrace \mathbf{S}_i \times \mathbf{S}_j\rbrace \\- \mathbf{\mathcal{B}}_\text{ext} . \mathbf{S}_i
\end{multline}
\noindent where, the indices i and j denote the lattice sites and the sum $\left<..\right>$  is over the nearest neighbours. The first term of Eq. (\ref{eq1}) represents the exchange interaction, and $J_\alpha$ are the exchange coefficients between the spins $\mathbf{S}_i$ and $\mathbf{S}_j$ at the atomic sites i and j. Here, we consider that the sublattice ABC for reference with $\alpha = (1-4)$ corresponds to the nearest-neighbor coupling of the lattice ABC as shown in Fig. \ref{fig1}. The parameters $J_\alpha > 0$ for antiferromagnetic Kagom\'e system; $K$ is the onsite anisotropy with easy axis along $\hat{e}_z$ direction. The third term of Eq. (\ref{eq1}) represents the contribution due to the total DMI energy. The parameter $\mathbf{D}_{ij}^\text{eff}$ incorporates the effective DMI vector calculated by performing the sum over all nearest neighbours between the spins obtained via the three-site model defined as 

\begin{equation}
\label{eq2}
 \mathbf{D}_{ij}^\text{eff} = \sum_k  \mathbf{\mathcal{D}}_{ijk} (\mathbf{R}_{ki},\mathbf{R}_{kj},\mathbf{R}_{ij})  
\end{equation}

\noindent where, $\mathbf{R}_{ki}$, $\mathbf{R}_{kj}$ are the distance vector from the site $k$ to the atomic sites $i$ and $j$ respectively.
The last term of Eq. (\ref{eq1}) corresponds to the harmonic magnetic field defined as 
\begin{equation}
\label{eq3}
    \mathbf{\mathcal{B}}_\text{ext} =  B_0 \sin (\Omega t) \hat{e}_z
\end{equation}
\noindent where, $B_0$ and $\Omega$ are the strength and frequency of the external magnetic field applied along $z$-direction. The time evolution of the spins in the i$^\text{th}$ sublattice in Kagom\'e system can be studied by using the Landau - Lifshitz - Gilbert (LLG) equation. 
\begin{equation}
\label{eq4}
    \frac{d\mathbf{S}_i}{dt} = -\gamma (\mathbf{S}_i \times \mathbf{H}^\text{eff}_i) + \alpha \mathbf{S}_i \times (\mathbf{S}_i \times \mathbf{H}^\text{eff}_i)
\end{equation}
\noindent where $\gamma$ is the gyromagnetic ratio and $\alpha$ is the Gilbert damping coefficient. The effective fields $\mathbf{H}^\text{eff}_i$ for the respective sub lattices can be obtained by $\mathbf{H}^\text{eff}_i = - \frac{\delta \mathcal{H}}{\delta \mathbf{S}_i}$. Using Eq. (\ref{eq1}) and (\ref{eq2}), we obtain nine nonlinear coupled first-order differential equations for the spins in the sublattice ABC. An explicit form of the coupled time evolution equation is given in Appendix (\ref{A1}).

\begin{figure*}[hbt]
\centerline
\centerline{ 
\includegraphics[scale=1.2]{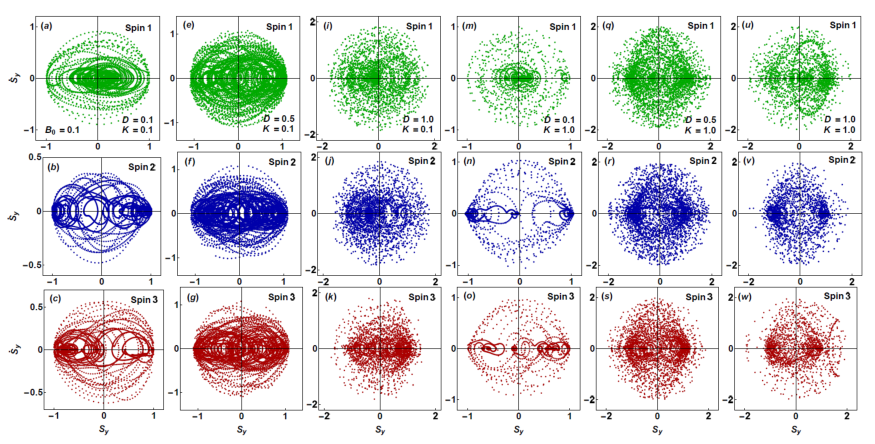}
}
\caption{The PSS for the Kagom\'e system having $(J_1, J_2, J_3, J_4) = (0.316, 0.285, 0.252, 0.145)$ for different values of $D$ and $K$  under magnetic field strength $B_0 = 0.1$ and frequency $\Omega = 1$. The plots in the top panel, middle and bottom panel are respectively correspond to Spin at the lattice site A (green), B (blue) and C (red) respectively.}
\label{fig2}
\end{figure*} 

\begin{figure}[hbt]
\centerline
\centerline{
\includegraphics[scale=0.6]{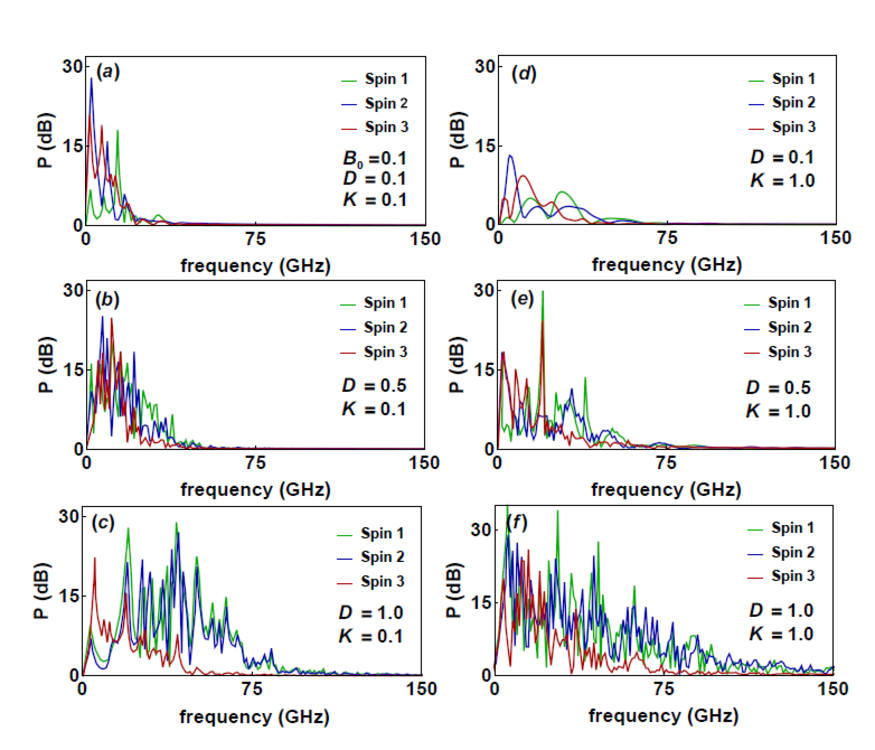}
}
\caption{The PS for Kagom\'e system for different values of $D$ and $K$ under magnetic field strength $B_0 = 0.1$ and frequency $\Omega = 1$. }
\label{fig3}
\end{figure}
\begin{figure*}[hbt]
\centerline
\centerline{ 
\includegraphics[scale=1.2]{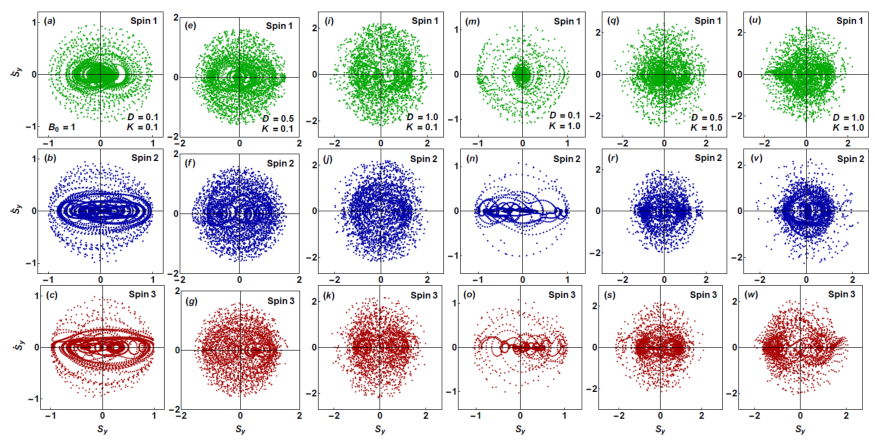}
}
\caption{The PSS for Kagom\'e system having $(J_1, J_2, J_3, J_4) = (0.316, 0.285, 0.252, 0.145)$ for different values of $D$ and $K$ under magnetic field strength $B_0 = 1.0$ and frequency $\Omega = 1$. The plots in the top panel, middle and bottom panel are respectively correspond to Spin at the lattice site A (green), B (blue) and C (red) respectively.}
\label{fig4}
\end{figure*}

\begin{figure}[hbt]
\centerline
\centerline{
\includegraphics[scale=0.6]{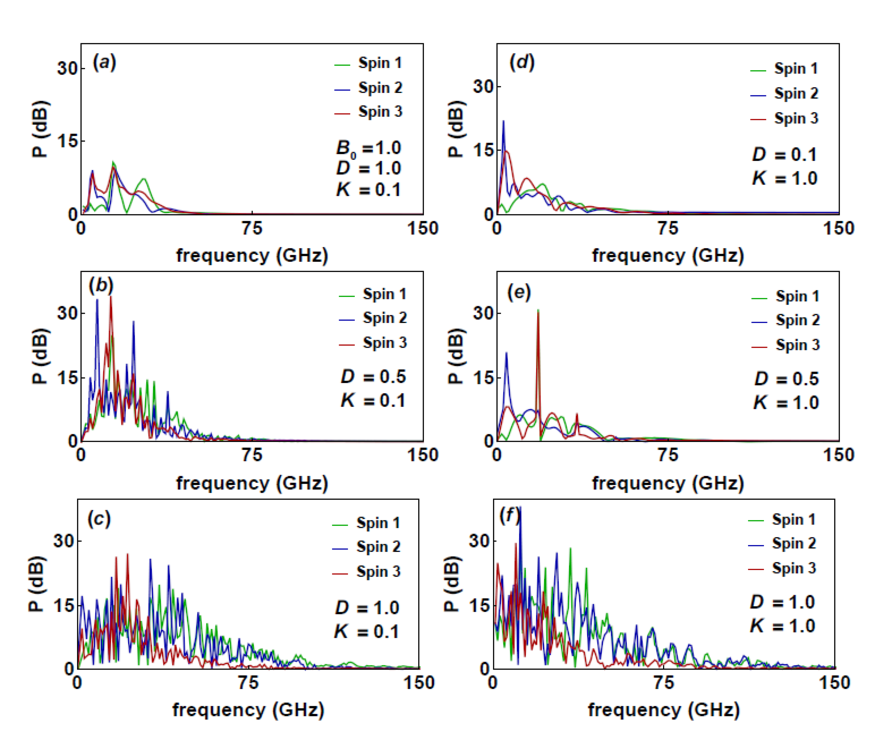}
}
\caption{The PS for Kagom\'e system for different values of $D$ and $K$ under magnetic field strength $B_0 = 1.0$ and frequency $\Omega = 1$. }
\label{fig5}
\end{figure}

\section{Simulations and Numerical Results}
This section is divided into two parts. The first part describes the dynamical behaviour of the Kagom\'e system using Poincar\'e Surface Section (PSS) and Power Sectra (PS). The subsequent part showcases the nature of oscillations using Magnetic resonance (MR) and Largest Lyapunov Exponents (LLE) and deliberates on our numerical findings.
For all our simulation we set the initial values $S_\text{A} = (1, 0, 0)$, $S_\text{B} = (0, 0, 1)$ and $S_\text{C} = (1, 0, 0)$. The simulations are performed by considering the exchange interaction values for the deformed Kagom\'e system Rb$_2$Cu$_3$SnF$_{12}$ with $(J_1, J_2, J_3, J_4) = (0.316, 0.285, 0.252, 0.145)$ \cite{khatami}. Additionally, we explore the impact of exchange coupling on spin dynamics in a subsequent section of this work. 

\subsection{Poincar\'e Surface Section (PSS) and Power Spectra (PS)}
Fig. \ref{fig2}  illustrates the PSS of the spins located at the lattice sites A, B, and C for weak magnetic field strength, $B_0 = 0.1$ with $\Omega = 1$ under weak and strong anisotropy conditions with different DMI strength. It is found that the system executes highly aperiodic and nonlinear oscillations for $(K, D) = (0.1, 0.1)$ as seen from Figs. \ref{fig2}(a) – \ref{fig2}(c). It is to be noted that the mutual exchange coupling dominates in this condition suppressing the impact of DMI and magnetic field. The corresponding PS in Fig. \ref{fig3}(a) indicates the presence of multiple low-frequency modes present in the oscillations. The presence of Kolmogorov-Arnold-Moser (KAM) islands indicates the non-chaotic nature of the oscillations.  With the increase in the strength of DMI to 0.5, the number of frequency modes increases and thereby enhances the aperiodicity of the oscillations as illustrated in the PSS Figs. \ref{fig2}(e) – \ref{fig2}(g) and corresponding PS in Fig. \ref{fig3}(b). For a system with strong DMI and weak anisotropy, i.e., $(K, D) = (0.1, 1.0)$, the PSS is found to display random blotch of points disappearing the KAM islands completely indicating the chaotic nature of the oscillations as observed from Figs. \ref{fig2}(i) – \ref{fig2}(k). The corresponding power spectra shown in Fig. \ref{fig3}(c), consist of a broad band of frequency modes indicating the non-linear and chaotic nature of oscillations. This is due to the reason that DMI dominates over the anisotropy and the magnetic field.  So, it is evident that the system undergoes a transition from quasi-periodic to strongly aperiodic oscillation with the increase in DMI strength for a system having low anisotropy and in the presence of a low magnetic field.

Figs. \ref{fig2}(m) – \ref{fig2}(w), represent the PSS for oscillations of the coupled spin oscillations in Kagom\'e system having strong uniaxial anisotropy $K = 1$. The coupled spin oscillations are found to be regular with the presence of multiple KAM islands for a system with $(K, D) = (1.0, 0.1)$ as seen from Figs. \ref{fig2}(m) – \ref{fig2}(o). The presence of only a few frequency modes in the corresponding PS in Fig. \ref{fig3}(d) indicates the regular nature of the oscillation. It is due to the stabilization of the spin oscillation due to the presence of strong anisotropy energy in the system. With the increase in $D$ to $0.5$, the KAM islands disappear with some splatter random points present in the PSS, indicating that the motion undergoes a transition from periodic to aperiodic oscillations as seen in Figs. \ref{fig2}(q) – \ref{fig2}(s). The corresponding PS in Fig. \ref{fig3}(e), consists of multiple frequency modes and indicates similar characteristics of the spin oscillations. A similar nature of oscillation in the PSS is also observed in Figs. \ref{fig2}(u) – \ref{fig2}(w) for a system with strong anisotropy and strong DMI strength, i.e., $(K, D) = (1.0, 1.0)$. The corresponding PS consist of more frequency bands as seen from Fig. \ref{fig3}(f). Thus, it is seen that the systems having strong anisotropy execute nearly chaotic behaviour for moderate and strong DMI strength. This is due to the reason that DMI introduces a torque that disturbs the regularity of spin oscillations. So, for strong anisotropic systems, the non-linearity is more evident as the non-linear oscillations are more prominent along the anisotropic direction. For low anisotropic systems, the nearly chaotic characteristics are seen only for the systems having strong DMI strengths.

\begin{figure*}[hbt]
\centerline
\centerline{
\includegraphics[scale = 0.45]{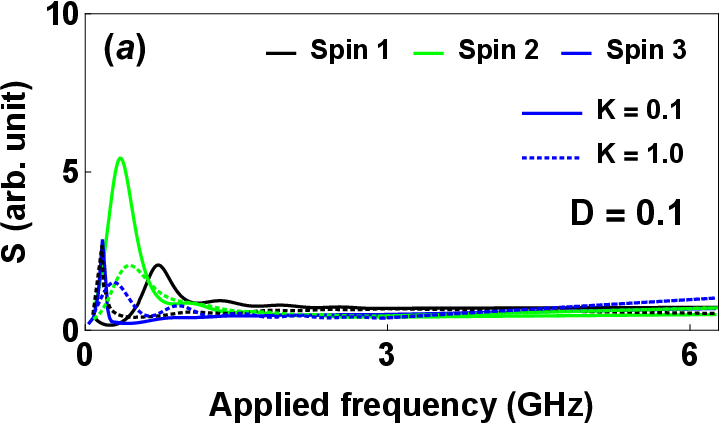}
\hspace{0.1cm}
\vspace{0.1cm}
\includegraphics[scale = 0.45]{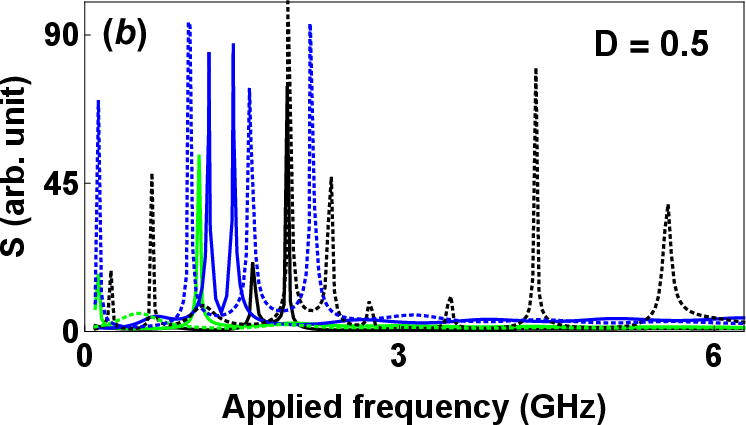}
\hspace{0.1cm}
\vspace{0.1cm}
\includegraphics[scale = 0.45]{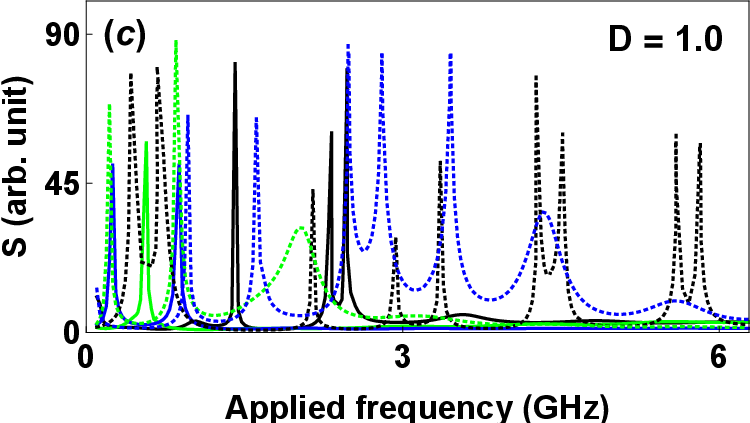}}
\caption{MR sprectra of Kagom\'e system for $B_0 = 1.0$ with DMI strength (a) $D = 0.1$, (b) $D = 0.5$ and (c) $D = 1.0$. The solid and dotted lines respectively correspond to $K = 0.1$ and $K = 1.0$. }
\label{fig6}
\end{figure*}

\begin{figure*}[hbt]
\centerline
\centerline{
\includegraphics[scale = 0.6]{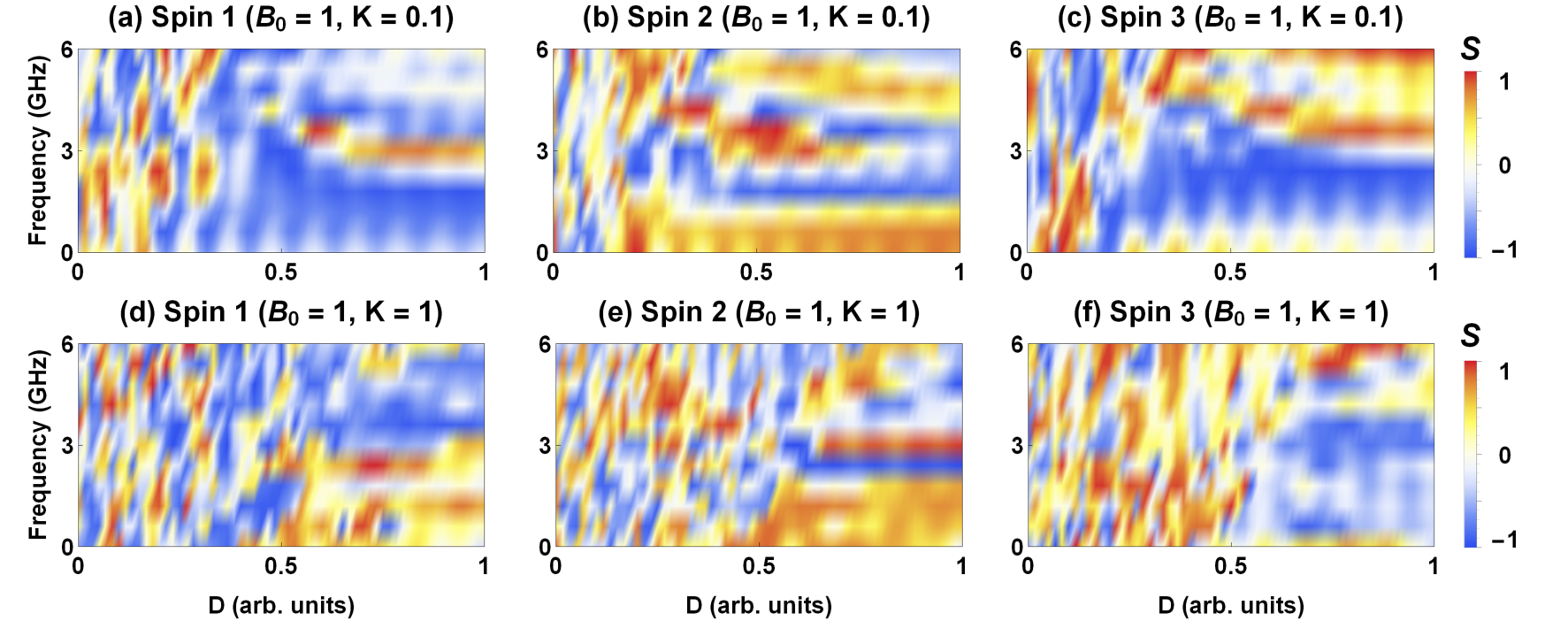}
\hspace{-0.1cm}
}
\caption{Density plot of MR with $D$ and frequency $\omega$ for $B_0 = 1$ with anisotropy energy $K = 0.1$ (top panel) and $K = 1.0$ (bottom panel).}
\label{fig7}
\end{figure*}
\begin{figure*}[hbt]
\centerline
\centerline{
\includegraphics[scale = 0.33]{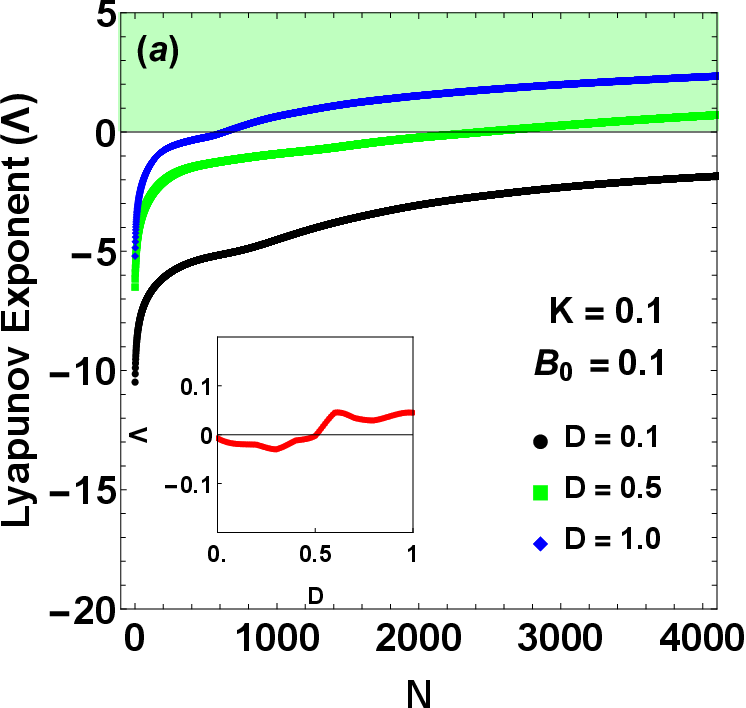}
\hspace{0.001mm}
\vspace{0.1cm}
\includegraphics[scale = 0.33]{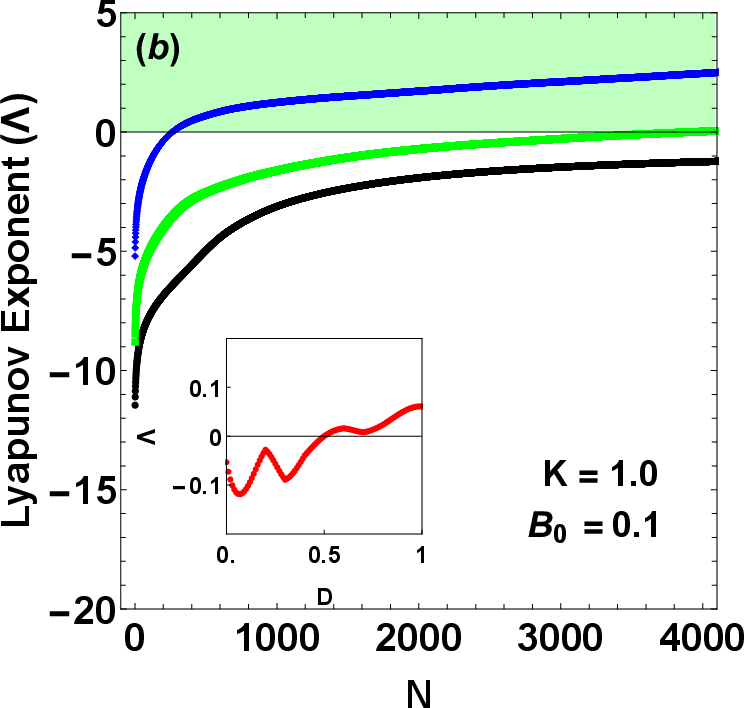}
\hspace{0.001mm}
\vspace{0.1cm}
\includegraphics[scale = 0.33]{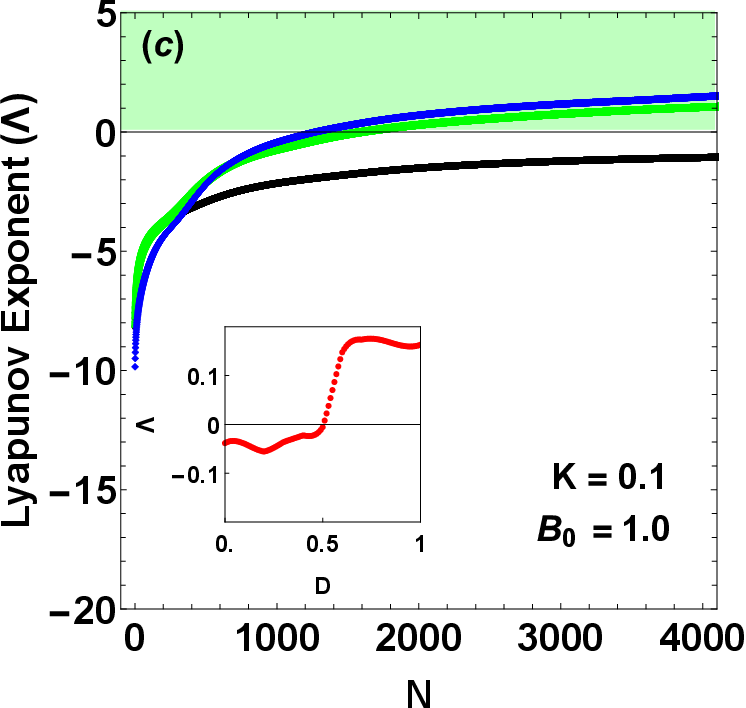}
\hspace{0.001mm}
\vspace{0.1cm}
\includegraphics[scale = 0.33]{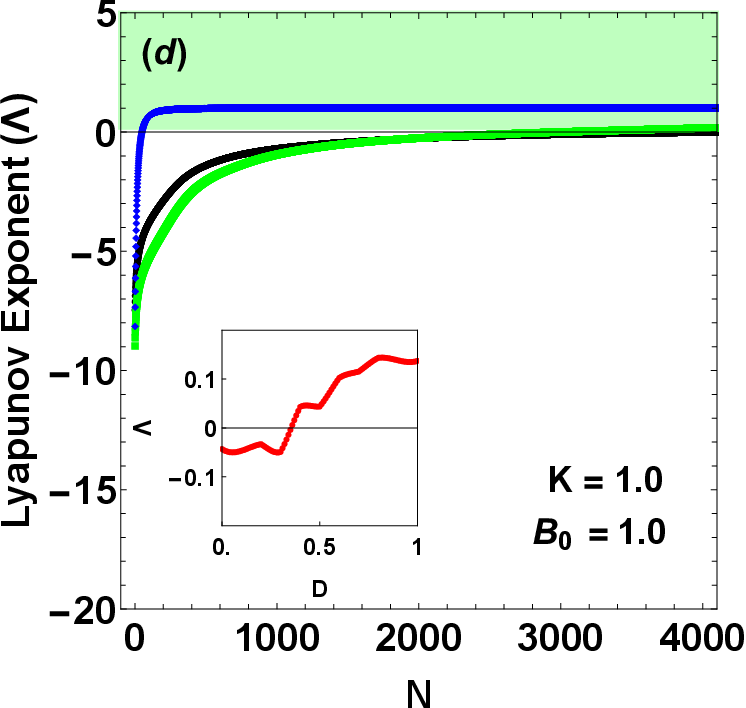}
\hspace{0.001mm}
}
\caption{Lyapunov exponents for Kagom\'e system for different choices of DMI strength with $\Omega = 1$ under (a) $B_0 = 0.1, K = 0.1$,
(b) $B_0 = 0.1, K = 1.0$, (c) $B_0 = 1.0, K = 0.1$ and (d) $B_0 = 1.0, K = 1.0$.}
\label{fig8}
\end{figure*}

To understand the impact of a strong magnetic field, the Poincar\'e surface of section (PSS) of the coupled spin system has been analyzed for a magnetic field strength of $B_0 = 1.0$ in Fig. \ref{fig4} under both weak and strong anisotropy conditions, with varying Dzyaloshinskii-Moriya interaction (DMI) strengths. The corresponding phase space (PS) is shown in Fig. \ref{fig5}. We observed that the nature of the oscillations is quasi-periodic for $(K, D) = (0.1, 0.1)$, as evidenced by the PSS in Figs. \ref{fig4}(a) – \ref{fig4}(c) and the PS in Fig. \ref{fig5}(a). This characteristic is similar to the system with a low magnetic field strength, as previously discussed. The PSS exhibits random points for $D \ge 0.5$, indicating aperiodic oscillations, as seen in Figs. \ref{fig4}(e) – \ref{fig4}(k) under low anisotropy conditions. This behavior is attributed to the increased magnetic field strength disrupting the spin oscillations, thus enhancing the aperiodicity of the oscillations, as evident from the multiple frequency nodes in Figs. \ref{fig5}(b) and \ref{fig5}(c). For materials with strong anisotropy, the system's characteristics in the PSS are similar to those of systems with strong anisotropy and low magnetic field conditions, as shown in Figs. \ref{fig2}(m) – \ref{fig2}(w). With an increase in anisotropy energy $(K = 1)$, the oscillations of the coupled spin system remain independent of the magnetic field strength, as demonstrated in Figs. \ref{fig4}(m) – \ref{fig4}(w) and Figs. \ref{fig2}(m) – \ref{fig2}(w). 

\subsection{Magnon dispersion and Magnetic Resonance (MR) in Kagom\'e system}
Over the years, Magnetic Resonance (MR) has been a standard tool for researchers to understand various couplings like magnetic anisotropy, exchange coupling, damping, and other key parameters that are essential for designing and gaining a deeper understanding of the magnetic structures and devices \cite{elkins, bakuzis, anisimov, yuan3, Chen91}. Moreover, it is a standard tool for probing spin wave excitation's and spin dynamics \cite{bauer, flovik, bertelli, crooker}.  Also, as already evident from Figs. \ref{fig2} –\ref{fig5}, the coupled spin oscillations of our system are strongly dependent on DMI strength, anisotropy, and applied magnetic field. Thus, to gain a better understanding of the nature of oscillation and its correlation with the applied magnetic field it is necessary to explore MR of our proposed system under weak and strong anisotropy conditions for different DMI strengths.

To obtain the magnon dispersion relation we consider the sub-lattice ABC of the Kagom\'e system defined by the order parameters $\mathbf{S}_\text{A}$, $\mathbf{S}_\text{B}$ and $\mathbf{S}_\text{C}$. In view of Eq. (\ref{eq4}), the equation of motion of the sub-lattice  ABC of the Kagom\'e system can be written as
\begin{align}
\label{eq14}
\dot{\mathbf{S}}_\text{A} &=-\gamma (\mathbf{S}_\text{A}\times \mathbf{H}^\text{eff}_\text{A}) + \alpha \mathbf{S}_\text{A} \times (\mathbf{S}_\text{A} \times \mathbf{H}^\text{eff}_\text{A})\\
\label{eq15}
\dot{\mathbf{S}}_\text{B} &=-\gamma (\mathbf{S}_\text{B}\times \mathbf{H}^\text{eff}_\text{B}) + \alpha \mathbf{S}_\text{B} \times (\mathbf{S}_\text{B} \times \mathbf{H}^\text{eff}_\text{B})\\
\label{eq16}
\dot{\mathbf{S}}_\text{C} &=-\gamma (\mathbf{S}_\text{C}\times \mathbf{H}^\text{eff}_\text{C}) + \alpha \mathbf{S}_\text{C} \times (\mathbf{S}_\text{C} \times \mathbf{H}^\text{eff}_\text{C})
\end{align}
where, $\mathbf{H}^\text{eff}_\text{A}$, $\mathbf{H}^\text{eff}_\text{B}$ and $\mathbf{H}^\text{eff}_\text{C}$ are the effective field of sub-lattice A, B and C respectively. In order to obtain the magnon dispersion we consider a spin wave of frequency $\omega$ and propagation vector $k$ defined as $\mathbf{S}_j = \mathbf{S}_j^0 + \delta\mathbf{S}_je^{\{i(\mathbf{k}.\mathbf{r} - \omega t)\}}$ with, $j = 1, 2, 3$ corresponding to the sub-lattice A, B and C respectively, with $\mathbf{S}_j^0$ being the ground state spin moment of the sub-lattice $j$. Here, we consider a small deviation perpendicular to $\mathbf{S}_j^0$ as $\delta\mathbf{S}_j$ is. The Eqs. (\ref{eq14}) - (\ref{eq16}) can be linearized by using Kittel's approach. In view of this the coefficient of $\delta\mathbf{S}_\text{A}$, $\delta\mathbf{S}_\text{B}$ and $\delta\mathbf{S}_\text{C}$ can be expressed in the form of a determinant as
 \begin{equation}
\label{eq17}
\begin{matrix}
\mathcal{S} = =\left(
\begin{array}{cccccc}
 -i \omega  & \mathcal{R}_1 &   D^z &   -J_1 &   -D^z &   -J_4 \\
 -\mathcal{R}_1 & -i \omega  &   J_1 &   D^z &   J_4 &   -D^z \\
   -D^z &   -J_1 & -i \omega  & \mathcal{R}_2 &   D^z &   -J_2 \\
   J_1 &   -D^z & -\mathcal{R}_2 & -i \omega  &   J_2 &   D^z \\
   D^z &   -J_4 &   -D^z &   -J_2 & -i \omega  & \mathcal{R}_3 \\
   J_4 &   D^z &   J_2 &   -D^z & -\mathcal{R}_3 & -i \omega  \\
\end{array}
\right)
\end{matrix}
\end{equation}
\\ 
where, $\mathcal{R}_1= -i \alpha  \omega -\gamma  (J_1-J_4-2K)$,  $\mathcal{R}_2 = -i \alpha  \omega +\gamma (J_1+J_2+2K)$ and $\mathcal{R}_3 = i \alpha  \omega +\gamma  (J_2-J_4-2K)$. The magnon dispersion relation can be obtained by solving the secular determinant $det(\mathcal{S}) = 0$ for the frequency $\omega$, can be expressed as
 \begin{equation}
\label{eq18}
\mathcal{A}_1\omega+\mathcal{A}_2\omega^2+\mathcal{A}_3\omega^3+\mathcal{A}_4\omega^4+\mathcal{A}_5\omega^5+\mathcal{A}_6\omega^6+ \xi = 0
\end{equation}
where, the $\mathcal{A}_i$ with $i = 1-6$ and $\xi$ are the coefficients whose explicit form is displayed in the Appendix (\ref{B1}).

The numerically simulated results of Eq. (\ref{eq18}) has been presented in Figs. \ref{fig6} and \ref{fig7} under weak and strong anisotropy conditions for different DMI strengths. We observed that under low DMI conditions, the system displays a single resonance peak as shown in Fig. \ref{fig6}(a) for low and high anisotropy conditions. This is due to the reason that the oscillations of the coupled system are nearly periodic in this condition. With the rise in DMI strength to $0.5$ and $1$, the system displays multipeaked resonance spectra as seen in Figs. \ref{fig6}(b) and \ref{fig6}(c). This is due to the reason that: (1) As the DMI strength increases, the oscillations of the coupled spin system undergo a transition from regular to aperiodic oscillations as observed earlier in Figs. \ref{fig2} –\ref{fig5} and (2) DMI can cause a change in the excitation of standing spin wave modes resulting in multipeaked MR spectra. The presence of multiple frequencies in the PS in this regime indicates the possibility of multiple resonance frequencies in the system. It is to be noted that systems with strong magnetic anisotropy can significantly influence the excitation of standing spin wave modes, resulting in the increase of MR peaks in the resonance spectra as depicted by the dotted lines in Figs. \ref{fig6}(b) and \ref{fig6}(c). Additionally, due to the aperiodic nature of the oscillations, it is essential to explore different initial conditions to understand the system's sensitive dependence.

\begin{figure*}[hbt]
\centerline
\centerline{
\includegraphics[scale = 0.55]{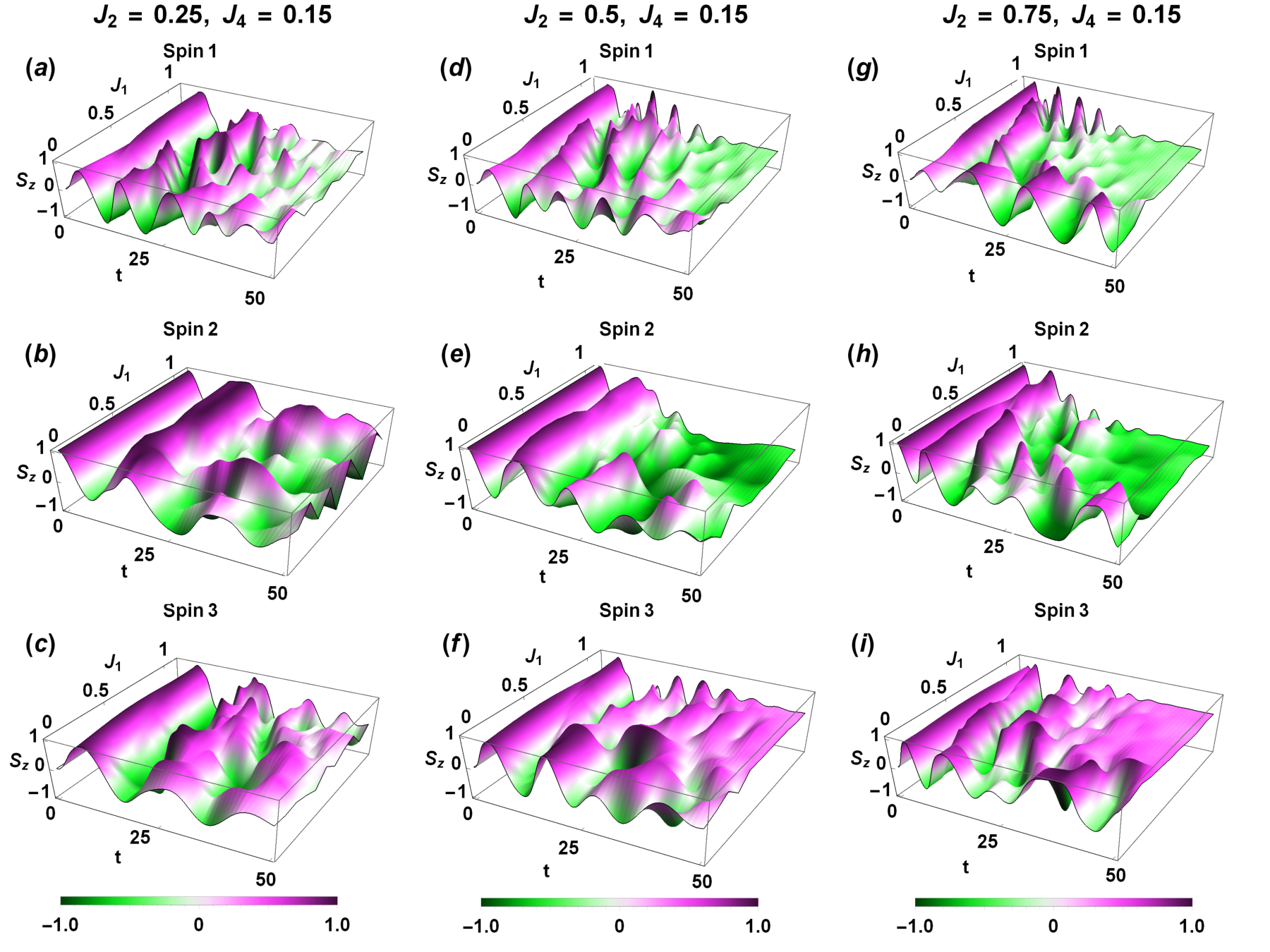}
\hspace{-0.1cm}
}
\caption{Variation of $S_z (t)$ with $t$ and $J_1$ considering $J_3 = 0.25$, $J_4 = 0.15$, $B_0 = 0.1$, $D = 0.5$, $K = 0.5$ and $\Omega = 1$. The plots in the left  panel (a)-(c) are for $J_2 = 0.25$, middle panel (d)-(f) are for $J_2 = 0.5$ and (g)-(i)  are for $J_2 = 0.25$. }
\label{fig9}
\end{figure*}

Figs. \ref{fig7} displays the density plot of the MR for different DMI and anisotropy strength for $B_0 = 1.0$. In this case, we set the initial conditions as $\{\textbf{S}_\text{A}, \textbf{S}_\text{B},  \textbf{S}_\text{C} \} = \{(1, 0, 0), (0, 1, 0), (1, 0, 0)\}$. Under low anisotropy conditions, the oscillations of the first spin ($\textbf{S}_\text{A}$) of the system are predominantly in the low DMI profile while it remains nearly independent with a maximum  $\sim 3$ GHz for moderate and high DMI strength. Moreover, the presence of multiple frequency bands at low DMI conditions indicates the sensitive dependence on initial conditions and the chaotic nature of oscillations. A similar characteristic is observed for ($\textbf{S}_\text{B}$) and ($\textbf{S}_\text{C}$) in the low DMI regime. However, for $\textbf{S}_\text{B}$ the MR spectra are predominated by broad peaks around low, mid, and high frequency for moderate and high DMI profiles. For $\textbf{S}_\text{C}$,  MR peaks are observed at moderate and high-frequency conditions for strong DMI strength as seen in Fig. \ref{fig7}(c). For systems having high anisotropy, multiple resonance frequencies are observed as the oscillations of the coupled spin system are highly aperiodic under low and moderate DMI strength while a quasi-periodic nature of resonance spectra is found for strong DMI systems as seen in Figs. \ref{fig7}(d) - \ref{fig7}(f).
Thus, from Fig. \ref{fig7}, we observe that the oscillations of the coupled spin system not only depend on the DMI strength but also have a significant dependence on the anisotropic energy of the system. Also, by understanding the MR spectroscopy, it may be possible to probe the complete magnetization dynamics of the proposed system can be utilized to fabricate microwave devices. 

\subsection{Largest Lyapunov Exponents $(\Lambda)$}
To comprehend the nature of spin oscillations of Kagom\'e spectra we analyze the time evolution of the order parameters in Figs. \ref{fig2} - \ref{fig5}. However, to gain insight into the stability of the system we have calculated the  Lyapunov Exponents (LE) \cite{sandri, wolff} of the system. The LEs are the indicators that measure the rate of convergence or divergence of the nearby trajectories. In the context of chaotic systems, a substantial increase in the rate of divergence is observed, leading to the emergence of positive values of LE. Conversely, for regular systems, the LE exhibits a negative value.

The LE can be obtained by computing the average sequence of distances $\Pi_i$ (where, $i = 0,1, .... N$) between the nearest neighbour trajectories over a finite time. The underlying concept involves initiating two closely situated points with initial displacements of $\mathcal{P}|_{(0,0)}$ and $\mathcal{Q}|_{(0,0)}$ at time $t=0$. Thus, we can express $\mathcal{Q}|_{(0,0)} = \mathcal{P}|_{(0,0)}+\bf{\Pi}_0$ initially at $t=0$, where $\bf{\Pi}_0$ signify the initial displacement between the two points. The points $\mathcal{P}|_{(0,0)}$ and $\mathcal{Q}|_{(0,0)}$ evolve after a time interval $\tau$, resulting in new positions $\mathcal{P}|_{(0,\tau)}$ and $\mathcal{Q}|_{(0,\tau)}$. So, we have $\mathcal{Q}|_{(0,\tau)} = \mathcal{P}|_{(0,\tau)} + \Pi_1$, where $\Pi_1 = |\Pi_1|$ is the new distance between the two points. Considering it to be the fresh reference point, we can write, $\mathcal{Q}|_{(1,0)} = \mathcal{P}|_{(1,0)} + \frac{\Pi_0}{\Pi_1} \mathbf{\Pi_1}$, with $\mathcal{P}|_{(1,0)} = \mathcal{P}|_{(0,\tau)}$. Following this iterative approach one can generate the second set of points ($\mathcal{P}|_{(1,\tau)}$, $\mathcal{Q}|_{(1,\tau)}$), where, $\mathcal{Q}|_{(1,\tau)} = \mathcal{P}|_{(1,\tau)} + \mathbf{\Pi}_2$, Resetting the reference points again to $\mathcal{Q}|_{(2,0)} = \mathcal{P}|_{(2,0)} +\frac{\Pi_0}{\Pi_2}\mathbf{\Pi_2}$, where, $\mathcal{P}|_{(2,0)} = \mathcal{P}|_{(1,\tau)}$ and continuing the iteration further one can obtain new set of points for a single time step of duration $\tau$. Repeating the steps for N times and generating a sequence of distances $\Pi_i =| \bf{\Pi}_i|$, where i = 1, 2,... N we obtain the finite time Largest Lyapunov Exponent (LLE), defined as \cite{acharjee5,liu}
\begin{equation}
\label{eq19}
\Lambda = \frac{1}{N \tau} \sum_{i = 1}^N \log_e\left(\frac{\bf{\Pi}_i}{\bf{\Pi}_0}\right)
\end{equation}
Here, $\Lambda$ is independent of $\tau$ as it is considered $\tau \ll 0$. For normal trajectories $\lim_{N\rightarrow\infty} \Lambda \leq 0$ while $\lim_{N\rightarrow\infty} \Lambda > 0$ for chaotic trajectories \cite{liu}.

The LLE of the Kagom\'e system corresponding to the PSS in Fig. \ref{fig2} and Fig. \ref{fig4}, is displayed in Fig. \ref{fig8}. The inset pictures corroborate the precise DMI value above which the system exhibits chaotic oscillations.  The spin oscillations are found to be mostly stable and non-chaotic for a system with weak anisotropy energy and weak DMI strength i.e., $(K, D) = (0.1, 0.1)$ in the presence of a weak harmonic field of frequency $\Omega = 1$. With the increase in DMi strength to $0.5$ the system displays highly unstable characteristics. For a system with strong DMI strength and weak anisotropy energy i.e., $(K, D) = (0.1, 1.0)$, the LLE is found to be positive suggesting that spin oscillations are strongly chaotic as seen from Fig. \ref{fig8}(a). The fig inset shows that the LLE, ($\Lambda < 0$) for $D \le 0.5$ indicate the normal to chaotic crossover region in this condition. A similar characteristic is also seen in Fig. \ref{fig8}(b) for $K = 1.0$ and $B_0 = 0.1$. In this case, the system also shows chaotic behaviour near the region $D \ge 0.5$.  Although the system's characteristics are identical, in this condition the system becomes strongly chaotic for $D \sim 0.5 $ and then execute unstable characteristics for $0.5 \le D < 0.7$. With the further increase in DMI strength, the system becomes strongly chaotic as seen from the inset of Fig. \ref{fig8}(b). The spin oscillations are stable for a system with weak anisotropy energy $K = 0.1$ and weak DMI strength $D = 0.1$ under strong magnetic field strength $B_0 = 1.0$. However, it becomes highly chaotic for $D > 0.5$ as seen from Fig. \ref{fig8}(c). It is to be noted that a stiff rise in the LLE is observed at $D \sim  0.5$ indicating a sudden transit from regular to chaotic regime in this condition. The nature of the oscillations is found to be drastically different in Fig. \ref{fig8}(d)  for a system with $(K, D) = (1.0, 1.0)$ with $B_0 = 1.0$. In this case, the LLE is positive for $D \ge 0.3$ indicating the early transit from regular to chaotic region in this regime.

\subsection{Coupled Spin Oscillations for different exchange strength}
While our primary analyses are confined to the exchange interaction of Rb$_2$Cu$_3$SnF$_{12}$, we are also interested in examining the impact of various exchange interactions on spin dynamics. In Fig. \ref{fig9}, we present the study of the z-component of spin for different Kagom\'e lattices, characterized by different values of $J_1$ and $J_2$. For this analysis, we set, $J_3 = 0.25$, $J_4 = 0.15$, $B_0 = 0.1$, $D = 0.5$, $K = 0.5$ and $\Omega = 1$ for our analysis. The oscillations of the system are found to be nearly periodic for low values of $J_1$ and die out over time.  A signature of aperiodic oscillations is noticed with the rise in $J_1$. However, the effect of damping torque is enhanced in this condition resulting in the spin oscillations at site A dying out too rapidly in this scenario as seen from Fig. \ref{fig9}(a).  For spins at the lattice site B and spin C, the oscillations are found to have low periodicity and are not much affected by the increase in $J_1$. Although some aperiodicity in spin oscillations is noticed, it is found to have periodicity and display weak damping with the increase in time $(t)$ as depicted in Fig. \ref{fig9}(b)  and Fig. \ref{fig9}(c). The rapid damping of the oscillations at the lattice site A is due to its coupling with the spin at the site B which tends to stabilize the oscillations. 

To explore the effect of exchange coupling between spins at sites B and C  on the coupled spin dynamics, we increased $J_2$ to $0.5$. The damping in the spin oscillations is reduced and a nonlinear characteristic is observed for low values of $J_1$ in this condition. However, significant damping is observed for higher values of $J_1$. It is to be noted that the frequency of oscillations also increases with the increase in the value of $J_2$ to $0.5$. A similar characteristic is also observed for spins at the lattice sites B and C, characterized by a significant damping noticed for higher values of $J_1$. This is due to the reason that with the increase in exchange interaction, the spin moments at sites A and B tend to align and hence enhance the damping. However, as spin at site B is connected to A and C via the exchange coupling $J_1$ and $J_2$, so for low values of $J_1$ and $J_2$ the damping is comparatively lower as compared to the higher values of it as observed from Figs. \ref{fig9}(d) – \ref{fig9}(f). However, the frequency of oscillations increases with the increase in the strength of $J_1$ for the spins at sites A and B. Although the characteristics are found to be similar in Figs. \ref{fig9}(g) – \ref{fig9}(i), the further increase of $J_2$ to $0.75$, a higher damping is observed. In summary, oscillations in the Kagom\'e system exhibit periodicity with low exchange interaction, moderate DMI strength, under moderate or strong anisotropy energy. Nonlinear characteristics emerge for the Kagom\'e systems with moderate exchange coupling, while strong exchange coupling results in highly damped oscillations under moderate DMI strength and anisotropy conditions.

\section{Conclusions}
In summary, we have investigated the spin dynamics of frustrated Kagom\'e lattice system, characterizing the dependence of the nonlinearity of the spin oscillations under different DMI strengths, exchange field, and anisotropy energy induced via harmonic magnetic field. We showed that when the strength of all the interacting parameters DMI, external field, exchange field, anisotropy and applied magnetic field are low, the oscillations are quasi-periodic as the exchange field dominates the spin oscillations. The increase in DMI strength adds to the aperiodicity thereby making the oscillations highly aperiodic. A strong anisotropy tends to make the oscillations periodic but an increase in DMI eventually makes the oscillations chaotic. It was also found that the strength of the external magnetic tends to destabilize the periodicity of oscillations for a Kagom\'e system with weak easy-axis anisotropy energy while the nature of the oscillations remains independent of the strength of the external magnetic field for a Kagom\'e system with strong anisotropy. In conclusion, our comprehensive analysis of spin dynamics in a Kagom\'e lattice system demonstrates a clear transition from quasi-periodic to chaotic oscillations as DMI strength increases. The study of magnon dispersion and MR spectra reveals multiple resonance peaks at higher DMI strengths, suggesting a complex interplay between spin wave excitation and system parameters. The results of our analysis shed light on the importance of understanding the inherent DMI and anisotropy of the Kagom\'e lattice structure during the fabrication process for different applications. Though, Kagom\'e lattice structure boasts of unique and exotic characteristics, the inherent DMI and anisotropy of the material can greatly impact its overall behaviour and performance. Another encouraging route for future study is to explore the different means by which the DMI and the anisotropy of the system can be controlled. As a concluding remark, this study provides valuable insights into the intricate dynamics of spin systems, with potential implications for designing advanced magnetic materials and devices. Such studies can add a new dimension to the Kagom\'e lattice structure from the research perspective as well as in potential technological applications.

\appendix
\section{Time evolution equations for the Spin at the sub-lattice ABC}
\label{A1}
The time evolution equation for the spin at the lattice site A, B and C are given by
\begin{widetext}
\begin{multline}
 \Dot{S}_\text{A}^{x}(t) = \Gamma [-D^z \{\alpha  S_\text{A}^x S_\text{A}^y (S_\text{B}^x+S_\text{C}^x)+S_\text{A}^z (S_\text{B}^x+S_\text{C}^x)
 +\alpha  \{(S_\text{A}^y)^2+(S_\text{A}^z)^2) (S_\text{B}^y+S_\text{C}^y)\}+D^y \{\alpha  ((S_\text{A}^y)^2+(S_\text{A}^z)^2) (S_\text{B}^z+S_\text{C}^z+S_\text{D}^z\\+S_\text{E}^z)-S_\text{A}^y (S_\text{B}^x+S_\text{C}^x)+\alpha  S_\text{A}^x S_\text{A}^z (S_\text{B}^x+S_\text{C}^x)\}
 +D^x \{S_\text{A}^y (\alpha  S_\text{A}^x (S_\text{B}^z+S_\text{C}^z+S_\text{D}^z
 +S_\text{E}^z)+S_\text{B}^y+S_\text{C}^y)+S_\text{A}^z (\alpha  S_\text{A}^x  \{-(S_\text{B}^y+S_\text{C}^y)\}\\+S_\text{B}^z+S_\text{C}^z+S_\text{D}^z+S_\text{E}^z)\}-(S_\text{A}^y-\alpha  S_\text{A}^x S_\text{A}^z) \{2 K_0 S_\text{A}^z-B_0 \sin ( \Omega t )
 +J_3 (S_\text{D}^z+S_\text{E}^z)\}+J_1 \{\alpha  S_\text{A}^x S_\text{A}^y S_\text{B}^y-S_\text{A}^y (\alpha  S_\text{A}^y S_\text{B}^x+S_\text{B}^z)\\+\alpha  S_\text{A}^x S_\text{A}^z S_\text{B}^z
-\alpha  (S_\text{A}^z)^2 S_\text{B}^x+S_\text{A}^z S_\text{B}^y\}+J_4 \{\alpha  S_\text{A}^x S_\text{A}^y S_\text{C}^y-S_\text{A}^y (\alpha  S_\text{A}^y S_\text{C}^x+S_\text{C}^z)
 +\alpha  S_\text{A}^x S_\text{A}^z S_\text{C}^z-\alpha  (S_\text{A}^z)^2 S_\text{C}^x+S_\text{A}^z S_\text{C}^y\}]
\end{multline}
\begin{multline}
 \Dot{S}_\text{A}^{y}(t) = \Gamma [D^z \{\alpha  S_\text{A}^x \{S_\text{A}^x (S_\text{B}^x+S_\text{C}^x)+S_\text{A}^y (S_\text{B}^y+S_\text{C}^y)\}+\alpha  (S_\text{A}^z){}^2 (S_\text{B}^x+S_\text{C}^x)-S_\text{A}^z (S_\text{B}^y+S_\text{C}^y)\}+D^y \{S_\text{A}^x (-\alpha  S_\text{A}^y (S_\text{B}^z+S_\text{C}^z+S_\text{D}^z+S_\text{E}^z)\\+S_\text{B}^x+S_\text{C}^x)+S_\text{A}^z (\alpha  S_\text{A}^y (S_\text{B}^x+S_\text{C}^x)+S_\text{B}^z+S_\text{C}^z+S_\text{D}^z+S_\text{E}^z)\}+D^x (-\{\alpha  ((S_\text{A}^x){}^2+(S_\text{A}^z){}^2) (S_\text{B}^z+S_\text{C}^z+S_\text{D}^z+S_\text{E}^z)+S_\text{A}^x (S_\text{B}^y+S_\text{C}^y)+\alpha  S_\text{A}^y S_\text{A}^z (S_\text{B}^y
 \\+S_\text{C}^y)\})
 +(S_\text{A}^x+\alpha  S_\text{A}^y S_\text{A}^z) \{2 K_0 S_\text{A}^z-B_0 \sin (\Omega t )+J_3 (S_\text{D}^z+S_\text{E}^z)\}+J_1 \{\alpha  (S_\text{A}^x S_\text{A}^y S_\text{B}^x-(S_\text{A}^x){}^2 S_\text{B}^y-(S_\text{A}^z){}^2 S_\text{B}^y+S_\text{A}^y S_\text{A}^z S_\text{B}^z)+S_\text{A}^x S_\text{B}^z\\-S_\text{A}^z S_\text{B}^x\}+J_4 \{\alpha  (S_\text{A}^x S_\text{A}^y S_\text{C}^x-(S_\text{A}^x){}^2 S_\text{C}^y-(S_\text{A}^z){}^2 S_\text{C}^y+S_\text{A}^y S_\text{A}^z S_\text{C}^z)+S_\text{A}^x S_\text{C}^z-S_\text{A}^z S_\text{C}^x\}]
\end{multline}
\begin{multline}
 \Dot{S}_\text{A}^{z}(t) = \Gamma [D^z \{S_\text{A}^x \{\alpha  S_\text{A}^z (S_\text{B}^y+S_\text{C}^y)+S_\text{B}^x+S_\text{C}^x\}+S_\text{A}^y \{-\alpha  S_\text{A}^z (S_\text{B}^x+S_\text{C}^x)+S_\text{B}^y+S_\text{C}^y\}\}-D^y \{\alpha  S_\text{A}^x \{S_\text{A}^z (S_\text{B}^z+S_\text{C}^z+S_\text{D}^z+S_\text{E}^z)+S_\text{A}^x (S_\text{B}^x\\+S_\text{C}^x)\}+S_\text{A}^y (S_\text{B}^z+S_\text{C}^z+S_\text{D}^z+S_\text{E}^z)+\alpha  (S_\text{A}^y){}^2 (S_\text{B}^x+S_\text{C}^x)\}+D^x \{\alpha  S_\text{A}^y \{S_\text{A}^z (S_\text{B}^z+S_\text{C}^z+S_\text{D}^z+S_\text{E}^z)+S_\text{A}^y (S_\text{B}^y+S_\text{C}^y)\}-S_\text{A}^x (S_\text{B}^z+S_\text{C}^z+S_\text{D}^z+S_\text{E}^z)
 \\+\alpha  (S_\text{A}^x){}^2 (S_\text{B}^y+S_\text{C}^y)\}+\alpha  ((S_\text{A}^x){}^2+(S_\text{A}^y){}^2) \{-2 K_0 S_\text{A}^z+B_0 \sin ( \Omega t)-J_3 (S_\text{D}^z+S_\text{E}^z)\}+J_1 \{\alpha  S_\text{A}^z (S_\text{A}^x S_\text{B}^x+S_\text{A}^y S_\text{B}^y)-\alpha  S_\text{B}^z ((S_\text{A}^x){}^2+(S_\text{A}^y){}^2)\\+S_\text{A}^x (-S_\text{B}^y)+S_\text{A}^y S_\text{B}^x\}+J_4 \{\alpha  S_\text{A}^z (S_\text{A}^x S_\text{C}^x+S_\text{A}^y S_\text{C}^y)-\alpha  S_\text{C}^z ((S_\text{A}^x){}^2+(S_\text{A}^y){}^2)+S_\text{A}^x (-S_\text{C}^y)+S_\text{A}^y S_\text{C}^x\}]
\end{multline}
\begin{multline}
 \Dot{S}_\text{B}^{x}(t) = \Gamma [D^z \{\alpha  S_\text{B}^y (S_\text{A}^y S_\text{B}^y-S_\text{B}^x S_\text{C}^x-S_\text{B}^y S_\text{C}^y)+\alpha  (S_\text{B}^z){}^2 (S_\text{A}^y-S_\text{C}^y)+S_\text{A}^x (\alpha  S_\text{B}^x S_\text{B}^y+S_\text{B}^z)-S_\text{B}^z S_\text{C}^x\}+D^y \{\alpha  S_\text{B}^z (-S_\text{A}^z S_\text{B}^z+S_\text{B}^x S_\text{C}^x+S_\text{B}^z S_\text{C}^z)\\+\alpha  (S_\text{B}^y){}^2 (S_\text{C}^z-S_\text{A}^z)+S_\text{A}^x (S_\text{B}^y-\alpha  S_\text{B}^x S_\text{B}^z)-S_\text{B}^y S_\text{C}^x\}+D^x \{\alpha  S_\text{B}^x (S_\text{A}^y S_\text{B}^z-S_\text{A}^z S_\text{B}^y-S_\text{B}^z S_\text{C}^y+S_\text{B}^y S_\text{C}^z)-S_\text{A}^y S_\text{B}^y-S_\text{A}^z S_\text{B}^z+S_\text{B}^y S_\text{C}^y+S_\text{B}^z S_\text{C}^z\}\\+J_1 \{S_\text{A}^y (\alpha  S_\text{B}^x S_\text{B}^y+S_\text{B}^z)+S_\text{A}^z (\alpha  S_\text{B}^x S_\text{B}^z-S_\text{B}^y)+\alpha  (-S_\text{A}^x) ((S_\text{B}^y){}^2+(S_\text{B}^z){}^2)\}+J_2 \{\alpha  S_\text{B}^x S_\text{B}^y S_\text{C}^y-S_\text{B}^y (\alpha  S_\text{B}^y S_\text{C}^x+S_\text{C}^z)+\alpha  S_\text{B}^x S_\text{B}^z S_\text{C}^z-\alpha  (S_\text{B}^z){}^2 S_\text{C}^x\\+S_\text{B}^z S_\text{C}^y\}+(S_\text{B}^y-\alpha  S_\text{B}^x S_\text{B}^z) (B_0 \sin (\Omega t)-2 K_0 S_\text{B}^z)]
\end{multline}
\begin{multline}
 \Dot{S}_\text{B}^{y}(t) = \Gamma [D^z \{\alpha  S_\text{B}^x (-S_\text{A}^x S_\text{B}^x+S_\text{B}^x S_\text{C}^x+S_\text{B}^y S_\text{C}^y)+\alpha  (S_\text{B}^z){}^2 (S_\text{C}^x-S_\text{A}^x)+S_\text{A}^y (S_\text{B}^z-\alpha  S_\text{B}^x S_\text{B}^y)-S_\text{B}^z S_\text{C}^y\}+D^y \{\alpha  S_\text{B}^y (S_\text{A}^z S_\text{B}^x-S_\text{B}^x S_\text{C}^z+S_\text{B}^z S_\text{C}^x)\\+S_\text{A}^x (-(S_\text{B}^x+\alpha  S_\text{B}^y S_\text{B}^z))-S_\text{A}^z S_\text{B}^z+S_\text{B}^x S_\text{C}^x+S_\text{B}^z S_\text{C}^z\}+D^x \{\alpha  \{S_\text{B}^y S_\text{B}^z (S_\text{A}^y-S_\text{C}^y)+S_\text{A}^z ((S_\text{B}^x){}^2+(S_\text{B}^z){}^2)-S_\text{C}^z ((S_\text{B}^x){}^2+(S_\text{B}^z){}^2)\}+S_\text{B}^x (S_\text{A}^y-S_\text{C}^y)\}\\+J_1 \{\alpha  S_\text{A}^x S_\text{B}^x S_\text{B}^y-\alpha  S_\text{A}^y ((S_\text{B}^x){}^2+(S_\text{B}^z){}^2)+S_\text{A}^z (S_\text{B}^x+\alpha  S_\text{B}^y S_\text{B}^z)-S_\text{A}^x S_\text{B}^z\}+J_2 \{S_\text{B}^x (\alpha  S_\text{B}^x (-S_\text{C}^y)+\alpha  S_\text{B}^y S_\text{C}^x+S_\text{C}^z)+S_\text{B}^z (\alpha  S_\text{B}^y S_\text{C}^z-S_\text{C}^x)\\+\alpha  (-(S_\text{B}^z){}^2) S_\text{C}^y\}+(S_\text{B}^x+\alpha  S_\text{B}^y S_\text{B}^z) (-(B_0 \sin (\Omega t)-2 K_0 S_\text{B}^z))]
\end{multline}
\begin{multline}
 \Dot{S}_\text{B}^{z}(t) = \Gamma [D^z \{\alpha  S_\text{B}^z (S_\text{A}^x S_\text{B}^y-S_\text{A}^y S_\text{B}^x-S_\text{B}^y S_\text{C}^x+S_\text{B}^x S_\text{C}^y)-S_\text{A}^x S_\text{B}^x-S_\text{A}^y S_\text{B}^y+S_\text{B}^x S_\text{C}^x+S_\text{B}^y S_\text{C}^y\}+D^y \{\alpha  (S_\text{A}^x-S_\text{C}^x) ((S_\text{B}^x){}^2+(S_\text{B}^y){}^2)\\+S_\text{A}^z (\alpha  S_\text{B}^x S_\text{B}^z+S_\text{B}^y)+\alpha  S_\text{B}^x (-S_\text{B}^z) S_\text{C}^z-S_\text{B}^y S_\text{C}^z\}+D^x \{\alpha  \{S_\text{B}^y (-S_\text{A}^z S_\text{B}^z+S_\text{B}^y S_\text{C}^y+S_\text{B}^z S_\text{C}^z)-S_\text{A}^y ((S_\text{B}^x){}^2+(S_\text{B}^y){}^2)+(S_\text{B}^x){}^2 S_\text{C}^y\}+S_\text{B}^x (S_\text{A}^z-S_\text{C}^z)\}\\+J_1 \{S_\text{A}^y (\alpha  S_\text{B}^y S_\text{B}^z-S_\text{B}^x)-\alpha  S_\text{A}^z ((S_\text{B}^x){}^2+(S_\text{B}^y){}^2)+S_\text{A}^x S_\text{B}^y+\alpha  S_\text{A}^x S_\text{B}^x S_\text{B}^z\}+J_2 \{\alpha  S_\text{B}^z (S_\text{B}^x S_\text{C}^x+S_\text{B}^y S_\text{C}^y)-\alpha  S_\text{C}^z ((S_\text{B}^x){}^2+(S_\text{B}^y){}^2)+S_\text{B}^x (-S_\text{C}^y)\\+S_\text{B}^y S_\text{C}^x\}+\alpha  ((S_\text{B}^x){}^2+(S_\text{B}^y){}^2) (B_0 \sin (\Omega t)-2 K_0 S_\text{B}^z)]
\end{multline}
\begin{multline}
 \Dot{S}_\text{C}^{x}(t) = \Gamma [D^z \{\alpha  S_\text{C}^y (S_\text{B}^y S_\text{C}^y-S_\text{C}^x S_\text{A}^x-S_\text{C}^y S_\text{A}^y)+\alpha  (S_\text{C}^z){}^2 (S_\text{B}^y-S_\text{A}^y)+S_\text{B}^x (\alpha  S_\text{C}^x S_\text{C}^y+S_\text{C}^z)-S_\text{C}^z S_\text{A}^x\}+D^y \{\alpha  S_\text{C}^z (-S_\text{B}^z S_\text{C}^z+S_\text{C}^x S_\text{A}^x+S_\text{C}^z S_\text{A}^z)\\+\alpha  (S_\text{C}^y){}^2 (S_\text{A}^z-S_\text{B}^z)+S_\text{B}^x (S_\text{C}^y-\alpha  S_\text{C}^x S_\text{C}^z)-S_\text{C}^y S_\text{A}^x\}+D^x \{\alpha  S_\text{C}^x (S_\text{B}^y S_\text{C}^z-S_\text{B}^z S_\text{C}^y-S_\text{C}^z S_\text{A}^y+S_\text{C}^y S_\text{A}^z)-S_\text{B}^y S_\text{C}^y-S_\text{B}^z S_\text{C}^z+S_\text{C}^y S_\text{A}^y+S_\text{C}^z S_\text{A}^z\}\\+J_1 \{S_\text{B}^y (\alpha  S_\text{C}^x S_\text{C}^y+S_\text{C}^z)+S_\text{B}^z (\alpha  S_\text{C}^x S_\text{C}^z-S_\text{C}^y)+\alpha  (-S_\text{B}^x) ((S_\text{C}^y){}^2+(S_\text{C}^z){}^2)\}+J_1 \{\alpha  S_\text{C}^x S_\text{C}^y S_\text{A}^y-S_\text{C}^y (\alpha  S_\text{C}^y S_\text{A}^x+S_\text{A}^z)+\alpha  S_\text{C}^x S_\text{C}^z S_\text{A}^z-\alpha  (S_\text{C}^z){}^2 S_\text{A}^x\\+S_\text{C}^z S_\text{A}^y\}+(S_\text{C}^y-\alpha  S_\text{C}^x S_\text{C}^z) (B_0 \sin (\Omega t)-2 K_0 S_\text{C}^z)]
\end{multline}
\begin{multline}
 \Dot{S}_\text{C}^{y}(t) = \Gamma [D^z \{\alpha  S_\text{C}^x (-S_\text{B}^x S_\text{C}^x+S_\text{C}^x S_\text{A}^x+S_\text{C}^y S_\text{A}^y)+\alpha  (S_\text{C}^z){}^2 (S_\text{A}^x-S_\text{B}^x)+S_\text{B}^y (S_\text{C}^z-\alpha  S_\text{C}^x S_\text{C}^y)-S_\text{C}^z S_\text{A}^y\}+D^y \{\alpha  S_\text{C}^y (S_\text{B}^z S_\text{C}^x-S_\text{C}^x S_\text{A}^z+S_\text{C}^z S_\text{A}^x)\\+S_\text{B}^x (-(S_\text{C}^x+\alpha  S_\text{C}^y S_\text{C}^z))-S_\text{B}^z S_\text{C}^z+S_\text{C}^x S_\text{A}^x+S_\text{C}^z S_\text{A}^z\}+D^x \{\alpha  \{S_\text{C}^y S_\text{C}^z (S_\text{B}^y-S_\text{A}^y)+S_\text{B}^z ((S_\text{C}^x){}^2+(S_\text{C}^z){}^2)-S_\text{A}^z ((S_\text{C}^x){}^2+(S_\text{C}^z){}^2)\}+S_\text{C}^x (S_\text{B}^y-S_\text{A}^y)\}\\+J_1 \{\alpha  S_\text{B}^x S_\text{C}^x S_\text{C}^y-\alpha  S_\text{B}^y ((S_\text{C}^x){}^2+(S_\text{C}^z){}^2)+S_\text{B}^z (S_\text{C}^x+\alpha  S_\text{C}^y S_\text{C}^z)-S_\text{B}^x S_\text{C}^z\}+J_1 \{S_\text{C}^x (\alpha  S_\text{C}^x (-S_\text{A}^y)+\alpha  S_\text{C}^y S_\text{A}^x+S_\text{A}^z)+S_\text{C}^z (\alpha  S_\text{C}^y S_\text{A}^z-S_\text{A}^x)\\+\alpha  (-(S_\text{C}^z){}^2) S_\text{A}^y\}+(S_\text{C}^x+\alpha  S_\text{C}^y S_\text{C}^z) (-(B_0 \sin (\Omega t)-2 K_0 S_\text{C}^z))]
\end{multline}
\begin{multline}
 \Dot{S}_\text{C}^{z}(t) = \Gamma [D^z \{\alpha  S_\text{C}^z (S_\text{B}^x S_\text{C}^y-S_\text{B}^y S_\text{C}^x-S_\text{C}^y S_\text{A}^x+S_\text{C}^x S_\text{A}^y)-S_\text{B}^x S_\text{C}^x-S_\text{B}^y S_\text{C}^y+S_\text{C}^x S_\text{A}^x+S_\text{C}^y S_\text{A}^y\}+D^y \{\alpha  (S_\text{B}^x-S_\text{A}^x) ((S_\text{C}^x){}^2+(S_\text{C}^y){}^2)\\+S_\text{B}^z (\alpha  S_\text{C}^x S_\text{C}^z+S_\text{C}^y)+\alpha  S_\text{C}^x (-S_\text{C}^z) S_\text{A}^z-S_\text{C}^y S_\text{A}^z\}+D^x \{\alpha  \{S_\text{C}^y (-S_\text{B}^z S_\text{C}^z+S_\text{C}^y S_\text{A}^y+S_\text{C}^z S_\text{A}^z)-S_\text{B}^y ((S_\text{C}^x){}^2+(S_\text{C}^y){}^2)+(S_\text{C}^x){}^2 S_\text{A}^y\}+S_\text{C}^x (S_\text{B}^z\\-S_\text{A}^z)\}+J_1 \{S_\text{B}^y (\alpha  S_\text{C}^y S_\text{C}^z-S_\text{C}^x)-\alpha  S_\text{B}^z ((S_\text{C}^x){}^2+(S_\text{C}^y){}^2)+S_\text{B}^x S_\text{C}^y+\alpha  S_\text{B}^x S_\text{C}^x S_\text{C}^z\}+J_1 \{\alpha  S_\text{C}^z (S_\text{C}^x S_\text{A}^x+S_\text{C}^y S_\text{A}^y)-\alpha  S_\text{A}^z ((S_\text{C}^x){}^2+(S_\text{C}^y){}^2)\\+S_\text{C}^x (-S_\text{A}^y)+S_\text{C}^y S_\text{A}^x\}+\alpha  ((S_\text{C}^x){}^2+(S_\text{C}^y){}^2) (B_0 \sin (\Omega t)-2 K_0 S_\text{C}^z)]
\end{multline}
\end{widetext}
where, $\Gamma = \frac{\gamma}{1+\alpha^2}$ and $\{D^x,D^y,D^z\}$ are the components of DMI vector.

\section{Explicit form of the coefficients $\mathcal{A}_i$ and $\xi$}
\label{B1}
An explicit form of the coefficients $\mathcal{A}_1, \mathcal{A}_2, \mathcal{A}_3, \mathcal{A}_4, \mathcal{A}_5, \mathcal{A}_6$ and $\xi$ are expressed below
\begin{equation}
    \mathcal{A}_1 = -1-3 \alpha ^2-3 \alpha ^4-\alpha ^6
\end{equation}
\begin{equation}
    \mathcal{A}_2 = -4 i \alpha \gamma  \left(1+\alpha ^2\right)^2   (J_4+3 K)
\end{equation}
\begin{multline}
        \mathcal{A}_3 = 2 (\alpha ^2+1) \gamma ^2 (\alpha ^2 ((D^z)^2-2 J_1^2+J_1 (J_4-J_2)+J_4 (J_2\\+4 J_4)+20 J_4 K+30 K^2)+3 (D^z)^2+2 J_1^2+J_1 J_2\\-J_1 J_4+2 J_2^2-J_2 J_4+2 J_4^2+4 J_4 K+6 K^2)
    \end{multline}
\begin{widetext}
    \begin{multline}
        \mathcal{A}_4 = -4 i \alpha  \gamma ^3 \{(D^z)^2 (\alpha ^2 (J_1-4 K)+J_1+2 J_2-2 J_4-8 K\}-J_1^2 \{\alpha ^2 (J_2-3 J_4-8 K)-3 J_2+J_4\}+\alpha ^2 J_1 \{J_2 (J_4\\+4 K)-2 J_4 (J_4+2 K)\}+J_1 J_2 (2 J_2-3 J_4)-2 (J_2^2 (J_4+2 K)+J_4^3+6 J_4^2 K+12 J_4 K^2+12 K^3)-2 \alpha ^2 \{J_4^2 (J_2+J_4)\\+2 J_4 K (J_2+4 J_4)+20 J_4 K^2+20 K^3)\}
    \end{multline}
    \begin{multline}
        \mathcal{A}_5 = \gamma ^4 [-(\alpha ^2+9) (D^z)^4+2 (D^z)^2 (2 (\alpha ^2-3) J_1^2+J_1 (\alpha ^2 (J_2+3 J_4+12 K)+J_2+3 J_4+4 K)-\alpha ^2 \{J_4 (J_2-2 J_4)+24 K^2\}\\+7 J_2 J_4+8 J_2 K-6 J_4^2-8 J_4 K-16 K^2)-4 (\alpha ^2+1) J_1^4-4 (\alpha ^2+1) J_1^3 (J_2-J_4)+J_1^2 (-\alpha ^2 (J_2^2+24 K (J_2-3 J_4)+2 J_2 J_4-15 J_4^2\\-96 K^2)-9 J_2^2+14 J_2 J_4+24 J_2 K-9 J_4^2-8 J_4 K)+2 J_1 (J_2^2 (\alpha ^2 J_4+J_4+8 K)+\alpha ^2 J_2 (J_4^2+12 J_4 K+24 K^2)-J_2 J_4 (7 J_4+12 K)\\+2 J_4^3-6 \alpha ^2 J_4 (J_4+2 K)^2)-9 J_2^2 J_4^2-\alpha ^2 \{J_2^2 J_4^2+12 J_2 J_4 (J_4+2 K)^2+4 (J_4^4+12 J_4^3 K+48 J_4^2 K^2+80 J_4 K^3+60 K^4)\}-16 J_2^2 J_4 K\\-16 J_2^2 K^2+4 J_2 J_4^3-4 J_4^4-16 J_4^3 K-48 J_4^2 K^2-64 J_4
 K^3-48 K^4]
    \end{multline}
     \begin{multline}
        \mathcal{A}_6 =  -4 i \alpha  \gamma ^5 \{(D^z)^2-2 J_1^2+J_1 (J_4-J_2)+J_4 (J_2+2 J_4)+8 J_4 K+12 K^2\} (-(D^z)^2 (J_1+J_4-K)+J_1^2 (J_2-J_4-2 K)-J_1 J_2 K\\+J_1 J_4 (J_4+K)+J_2 J_4^2+J_4 K (J_2+2 J_4)+4 J_4 K^2+4 K^3)
    \end{multline}
    \begin{multline}
        \xi=  4 \gamma ^6 [-(D^z)^2 (J_1+J_4-K)+J_1^2 (J_2-J_4-2 K)-J_1 J_2 K+J_1 J_4 (J_4+K)+J_2 J_4^2+J_4 K (J_2+2 J_4)+4 J_4 K^2+4 K^3]^2
    \end{multline}
\end{widetext}

\end{document}